\documentclass[10pt,english,showpacs,letterpaper,prl,amsfonts,amssymb,oneside,balancelastpage,notitlepage,twocolumn,preprint]{revtex4}
\usepackage[T1]{fontenc}
\usepackage[latin1]{inputenc}
\usepackage{amsmath}
\usepackage{graphicx}
\usepackage{amssymb}

\makeatletter

\usepackage{amsfonts}



\usepackage{babel}
\makeatother
\begin{document}

\title{Nonlinear optical Galton board}

\author{C. Navarrete-Benlloch$^{1,2}$, A. Pérez$^{1}$ and Eugenio Roldán$^{2}$}

\affiliation{$^{1}$Departament de Física Teòrica and IFIC, Universitat de València-CSIC,
Dr. Moliner 50, 46100--Burjassot, Spain}

\affiliation{$^{2}$Departament d'Òptica, Universitat de València, Dr. Moliner
50, 46100--Burjassot, Spain}

\begin{abstract}
We generalize the concept of optical Galton board (OGB), first proposed
by Bouwmeester et al. {[}Phys. Rev. A \textbf{61}, 013410 (2000)],
by introducing the possibility of nonlinear self--phase modulation
on the wavefunction during the walker evolution. If the original Galton
board illustrates classical diffusion, the OGB, which can be understood
as a grid of Landau--Zener crossings, illustrates the influence of
interference on diffusion, and is closely connected with the quantum
walk. Our nonlinear generalization of the OGB shows new phenomena,
the most striking of which is the formation of non-dispersive pulses
in the field distribution (soliton--like structures). These exhibit
a variety of dynamical behaviors, including ballistic motion, dynamical
localization, non--elastic collisions and chaotic behavior, in the
sense that the dynamics is very sensitive to the nonlinearity strength.
\end{abstract}


\maketitle

\section{Introduction.}

The Galton board, or quincunx, is a matrix of regularly spaced pegs
fixed to a board through which pellets fall impulsed by gravity. The
final distribution of pellets' locations at the bottom of the device
follows the binomial distribution, and thus the Galton board constitutes
a realization of the random walk. The importance of random walks does
not need to be emphasized here, as their presence is ubiquitous in
science. They are important, in particular, as a tool in classical
computation (the best known algorithms for solving some particular
problems are based on their use \cite{Motwani}). For sure, this is
one of the main reasons behind the present interest on the quantum
counterpart of random walks, the so--called quantum random walks \cite{Aharonov}
or, more appropriately, quantum walks (QWs). Moreover, from a fundamental
point of view the study of quantum counterparts of important classical
phenomena, and viceversa, is of obvious interest.

The QW has been introduced from several different perspectives: In
the seminal papers (in 1993 Aharonov et al. \cite{Aharonov} introduced
the QW as a generalization of the random walk, and in 1996 Meyer \cite{Meyer}
introduced it as a nontrivial quantum cellular automaton) the computational
aspects were not stressed, but later Watrous \cite{Watrous} independently
introduced QWs from a quantum algorithmic point of view, and other
versions of the QW (the so--called continuous QW, that can be viewed
as a quantum generalization of the Markov chain) were also proposed
\cite{Farhi98}. Today there is a considerable amount of papers devoted
to QWs, and we refer the reader to existing reviews \cite{reviews,Kendon}.

Not only one can think of quantum versions of the random walk, one
can also think of \textit{wave} \cite{Bouwmeester} as well as \textit{quantum}
versions of the Galton board \cite{Sanders}. The difference between
the wave and quantum versions lies in that, in the wave version, it
is classical waves (e.g. optical waves) what are used, while the quantum
version uses amplitude probability waves (wavefunctions). The so--called
optical Galton board (OGB) was first proposed by Bouwmeester et al.
\cite{Bouwmeester}, and was introduced as a grid of (optical) Landau--Zener
crossings. Bouwmeester et al. showed, both theoretically and experimentally,
the existence of spectral diffusion, as well as dynamical localization
in their particular proposal for an OGB. As for the quantum version,
the quantum quincunx of Ref. \cite{Sanders} is a quantum--optical
proposal for the implementation of the QW.

Although classical waves and wavefunctions are different in a deep
sense, a very interesting point raised by the OGB is that it can be
understood, to some extent, as an optical realization of the QW \cite{Knight(b),Knight,Wojcik}.
It is convenient to stress that there are small differences between
the OGB of \cite{Bouwmeester} and the QW, but as it was shown in
\cite{Knight(b)}, the OGB reduces to a QW with an appropriate parameter
setting of the device. Moreover, Wojcik et al. \cite{Wojcik} suggested
that their generalization of the QW (consisting in the introduction
of some additional position-- dependent phase changes of the walker,
see also \cite{Buerschaper,Romanelli,Banyuls}) qualitatively describes
the OGB of \cite{Bouwmeester}, as it reproduces the observed dynamical
localization. These generalizations of the QW have shown unsuspected
connections of the QW with Anderson localization \cite{Romanelli}
and quantum chaos \cite{Buerschaper,Wojcik}.

Here we propose a nonlinear generalization of the OGB (NLOGB), consisting
on the introduction of nonlinearity in the evolution of the walker.
Given the connection between the OGB and the QW mentioned above, one
could say that we are proposing a nonlinear generalization of the
QW (the nonlinear QW). However, as we discuss below, our proposal
makes full sense only from a classical perspective, and thus our preference
for the name NLOGB (nonlinear optical Galton board). As expected,
the nonlinearity deeply modifies the QW dynamics, giving rise to new
and interesting phenomena which we investigate in some detail.

After this introduction, the rest of the article is organized as follows:
In Section II we briefly review the QW, as we will use its formalism
for introducing the NLOGB; in Section III we introduce the NLOGB as
a nonlinear QW; in Section IV we describe the formation of solitonic
structures; in Section V we analyze the dynamics of the system describing
the different phase transitions we have observed; and in Section VI
we give our main conclusions.

\section{The coined quantum walk.}

Here we deal with the coined, discrete QW, in one dimension. This
process is better introduced as a quantum generalization of the random
walk: In the random walk the walker moves to the right or to the left,
depending on the output of a random process (e.g. tossing a coin);
then the QW mimics the random walk in the existence of a conditional
displacement that depends on the state of the coin, but differs from
the QW in the fact that the coin is not a binary random process but
a qubit. As the qubit can be in a superposition state, the walker
can \textit{move simultaneously}, say, in the two opposite directions.
In order to make the dynamics nontrivial \cite{Meyer}, the coin state
must be changed (the analog of tossing the classical coin) after each
walk step, what is accomplished by the application of a suitable unitary
transformation on the qubit. The main feature of the QW, as opposed
to the random walk, is that the diffusion of the particle is much
faster (in the absence of decoherence \cite{Kendon}): While in the
random walk the width of the probability distribution of the walker
position grows as the square root of the number of steps, it grows
linearly with the number of steps in the QW. Moreover, the probability
distributions have a very different shape (gaussian in the random
walk, and resembling the Airy function in the QW). Let us now introduce
formally the QW.

The standard coined QW corresponds to the discrete time evolution,
on a one-dimensional lattice, of a quantum system (the walker), coupled
to a bidimensional system (the coin), under repeated application of
a pair of discrete linear operators. Let $\mathcal{H}_{W}$ be the
Hilbert space of the walker, with $\left\{ \left\vert m\right\rangle ,m\in\mathbb{Z}\right\} $
a basis of $\mathcal{H}_{W}$; and let $\mathcal{H}_{C}$ be the Hilbert
space of the coin, with basis $\left\{ \left\vert u\right\rangle ,\left\vert d\right\rangle \right\} $.
The state of the total system belongs to the space $\mathcal{H}=\mathcal{H}_{C}\otimes\mathcal{H}_{W}$
and, at time $t$, can be expressed as \begin{equation}
\left\vert \psi\left(t\right)\right\rangle =\sum_{m}\left[u_{m,t}\left\vert u,m\right\rangle +d_{m,t}\left\vert d,m\right\rangle \right].\end{equation}
 The connection between states in consecutive times is made by an
unitary linear evolution operator $\hat{U}$, which can be written
as $\hat{U}=\hat{U}_{d}\hat{U}_{c}$, i.e., $\left\vert \psi(t)\right\rangle =\hat{U}_{d}\hat{U}_{c}\left\vert \psi(t-1)\right\rangle $.
Here, $\hat{U}_{c}=\hat{C}\otimes\hat{I}$ is the \char`\"{}coin toss
operator\char`\"{} with $\hat{C}\in SU(2)$, typically chosen as the
Hadamard transformation \begin{equation}
\hat{C}=\frac{1}{\sqrt{2}}\left(\left\vert u\right\rangle \left\langle u\right\vert -\left\vert d\right\rangle \left\langle d\right\vert +\left\vert u\right\rangle \left\langle d\right\vert +\left\vert d\right\rangle \left\langle u\right\vert \right),\end{equation}
 and \begin{equation}
\hat{U}_{d}={\displaystyle \sum_{m}}\left(\left\vert u,m+1\right\rangle \left\langle u,m\right\vert +\left\vert d,m-1\right\rangle \left\langle d,m\right\vert \right),\end{equation}
 is the \char`\"{}conditional displacement operator\char`\"{}, which
moves the walker one position to the right or to the left, depending
on whether the coin state is $u$ or $d$, respectively. The main
quantity related to the walk is the probability distribution function
of the walker along the lattice, calculated as $P_{m}(t)=\left\vert u_{m,t}\right\vert ^{2}+\left\vert d_{m,t}\right\vert ^{2}\equiv P_{m}^{u}(t)+P_{m}^{d}(t)$. 

We have already commented that the QW can be classically simulated.
In order to make things concrete, we consider the scheme depicted
in Fig. 1, which represents an optical cavity. A quasi-monochromatic
light field enters the cavity through a partially reflecting mirror.
When this field reaches the beam--splitter (BS in Fig. 1), it can
follow two different paths, upper and lower in the figure. These two
paths play the role of the qubit (which, in this case, would be better
called a cebit, following the terminology introduced in \cite{Spreeuw}),
and the beam--splitter implements the unitary transformation \cite{Spreeuw}
(the \char`\"{}coin toss\char`\"{} operator). Then in the lower (upper)\ path,
the field frequency, which plays the role of the walker in this optical
implementation, is increased (decreased) in a fixed amount $\Delta\omega$
by means of appropriately tuned electrooptic modulators. This is the
first step of the QW. Then, the cavity mirrors reflect the light back
to the beam--splitter and a new step of the QW is implemented, and
so on and so forth.

\begin{figure}
\includegraphics[scale=0.45]{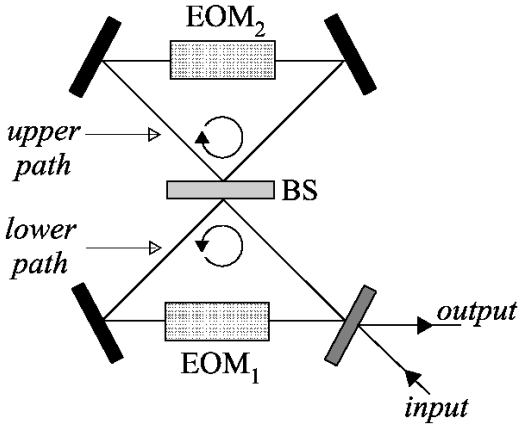}

\caption{Optical cavity for the implementation of the OGB and the NLOGB. EOM$_{1}$
and EOM$_{2}$ are two electrooptic modulators which are tuned for
incrementing (decreasing)\ the field frequency in $\Delta\omega$.
BS is a beam--splitter, and the cavity is constituted by four mirrors,
one of which is partially reflecting and serves as input/output port.
For implementing the OGB, the upper and lower paths must be a linear
optical medium, which must be replaced by a nonlinear optical medium
(such as, e.g., an optical fiber) for implementing the NLOGB.}
\end{figure}

In this case, the QW occurs in the frequency distribution of the output
field, with the intensity of each frequency component playing the
role of the probability of finding the walker at a given position,
i.e., $P_{m}^{u}(t)$ and $P_{m}^{d}(t)$ are spectral intensities
in this classical--wave context, and not probabilities. In other words,
after $m$ cavity roundtrips, the spectrum of the output field exhibits
the probability distribution of the QW. This is one of the schemes
proposed in \cite{Knight(b)} for the optical (classical) implementation
of the QW, where also the connection between the OGB of Ref. \cite{Bouwmeester}
and the QW\ is given, and we refer the reader to that paper for full
details on this type of classical (optical) implementation of the
QW. Let us emphasize that this scheme constitutes a realization of
the optical Galton board.

What this classical implementation of the QW (and others \cite{Hillery,Jeong,Roldan,Do05})
suggests is that interference, and not entanglement, is the responsible
of the QW characteristics. Entanglement would manifest in QWs in more
than two dimensions, in the amount of classical resources needed for
its implementation, as compared with a true quantum implementation,
as already discussed in \cite{Knight(b)}. This does not mean that
there is nothing quantum in the QW: It is the different physical meaning
of $P_{m}(t)$ (in a true quantum system, the probability distribution
can be reconstructed only after a large enough number of measurements,
while in the classical simulation the analog of the probability distribution
corresponds to the field spectrum and can be seen completely at each
walk step). The effect of decoherence could be different in classical
and quantum implementations \cite{Kendon}. But, at least in the QW
on the line, the quantum nature seems not to manifest, as it can be
successfully simulated by classical means. See \cite{Kendon05} for
a discussion on this topic.

\section{Introducing the nonlinear optical Galton board.}

The optical cavity scheme of Fig. 1 serves us to introduce the nonlinear
optical Galton board (NLOGB). It suffices to assume that light acquires
some intensity-dependent phase while traveling through the upper and
lower paths, i.e., that these paths are not made by a linear medium
(vacuum), but with a nonlinear medium (e.g. a Kerr medium, like an
optical fiber or similar). This is very easily taken into account
with the QW\ formalism introduced in the previous section that we
will continue to use here: We only need to introduce one more operator
describing the acquisition of the intensity-dependent (nonlinear)
phase due to propagation in the Kerr medium, i.e., we have to generalize
the unitary operator defined above in the following way: \begin{gather}
\hat{U}(t)=\hat{U}_{d}\hat{U}_{c}\hat{U}_{nl}(t-1),\label{Unl1}\\
\hat{U}_{nl}(t)=\underset{c=u,d}{\sum}\underset{m}{\sum}e^{iF_{c}\left(m,t\right)}\left\vert c,m\right\rangle \left\langle c,m\right\vert ,\label{Unl2}\end{gather}
 where $F_{c}\left(m,t\right)$ $\left(c=u,d\right)$ is an arbitrary
function of the probabilities (or intensities, in a classical context)
$P_{m}^{u}(t)$ and $P_{m}^{d}(t)$ \cite{nota}. Notice that the
role of $\hat{U}_{nl}(t)$ is to add a nonlinear (probability dependent)
phase to each of the spinor components. With the above formulation,
the standard QW is obviously recovered when $F_{u}=F_{d}=0$, and
the generalized QWs of \cite{Romanelli} and \cite{Buerschaper,Wojcik}
are recovered when $F_{u}=F_{d}=m^{2}\phi_{0}$ and $F_{u}=F_{d}=m\phi_{0}$,
respectively, with $\phi_{0}$ a constant phase. We see that a physical
system like the one represented in Fig. 1 allows to implement a number
of interesting generalizations of the QW in a relatively simple way.
Let us emphasize that the OGB of Bouwmeester et al. \cite{Bouwmeester}
is very close to what we are commenting \cite{Knight(b)}.

In this article we shall consider one of the simplest forms for (\ref{Unl2})
by choosing $F_{c}\left(m,t\right)=2\pi\alpha\left\vert c_{m,t}\right\vert ^{2}$
($c=u,d$), i.e., we assume that the nonlinear phase gained between
two QW steps is due to a Kerr-type nonlinearity that acts separately
on the two coin states ($u$ and $d$) and has a strength $\alpha$.
The recursive evolution equations for the probability amplitudes can
be easily derived from $\left\vert \psi(t+1)\right\rangle =\hat{U}\left(t+1\right)\left\vert \psi(t)\right\rangle $,
yielding\begin{align}
u_{m,t+1} & =\frac{1}{\sqrt{2}}u_{m-1,t}e^{i2\pi\alpha\left\vert u_{m-1,t}\right\vert ^{2}}\label{rec1}\\
 & +\frac{1}{\sqrt{2}}d_{m-1,t}e^{i2\pi\alpha\left\vert d_{m-1,t}\right\vert ^{2}},\nonumber \\
d_{m,t+1} & =\frac{1}{\sqrt{2}}u_{m+1,t}e^{i2\pi\alpha\left\vert u_{m+1,t}\right\vert ^{2}}\text{ }\label{rec2}\\
 & -\frac{1}{\sqrt{2}}d_{m+1,t}e^{i2\pi\alpha\left\vert d_{m+1,t}\right\vert ^{2}}.\nonumber \end{align}

As we show below, the nonlinearity just introduced deeply modifies
the behavior of the probability distribution $P_{m}(t)$. For this
purpose, we perform a numerical study of Eqs. (\ref{rec1},\ref{rec2})
for different values of $\alpha$. We shall consider $\alpha>0$ for
definiteness, since from Eqs. (\ref{rec1},\ref{rec2}) one easily
sees that choosing a positive $\alpha$, say $\alpha=\alpha_{0}$,
with some initial conditions $\left(u_{m,0};d_{m,0}\right)$ is equivalent
to taking $\alpha=-\alpha_{0}$ and complex-conjugated initial conditions
$\left(u_{m,0}^{\ast};d_{m,0}^{\ast}\right)$. Moreover, we shall
adopt, unless otherwise specified, symmetrical initial conditions
localized at the origin, i.e., $u_{m,0}=\delta_{m0}/\sqrt{2}$ and
$d_{m,0}=i\delta_{m0}/\sqrt{2}$.

From a classical (wave) viewpoint, the above process is a nonlinear
optical Galton Board (NLOGB) and can be implemented with the same
device we have commented in the previous section, provided that the
two optical beams propagate in a Kerr medium (e.g., an optical fiber),
as this nonlinear propagation exactly corresponds to what $\hat{U}_{nl}$
represents. From a quantum viewpoint the implementation of $\hat{U}_{nl}$
is probably impossible because of the linearity of the Schrödinger
equation. It is clear from now that the process we are proposing makes
full sense only as a nonlinear OGB, and will find conceptual difficulties
as a nonlinear QW.

In spite of the difficulties when speaking of a nonlinear QW, one
should keep in mind that nonlinearities can be introduced in quantum
systems through a clever use of measurement \cite{Knill,Sanaka},
what keeps open the possibility of implementing the proposed NLQW.
Another, more realistic possibility concerns using systems described
by nonlinear \textit{effective Hamiltonians}, as Bose--Einstein condensation,
where QWs could be implemented \cite{Chandrashekar}, or superconducting
devices, just to mention a couple of potential candidates. But these
appear as remote possibilities, as compared with the immediacy of
an optical implementation in an optical device similar to that already
used by Bouwmeester et al. \cite{Bouwmeester}.

\section{Formation of soliton--like structures.}

In Fig. 2 we represent the evolved probability distributions $P_{m}(t)$
for $\alpha=0$ (i.e., the standard QW) and $\alpha=0.4$. When $\alpha=0$,
we observe the typical QW behavior \cite{reviews}: $P_{m}(t)$ exhibits
two peaks at the borders of the distribution, whose tails decay in
the central zone, and whose maximum value monotonically decreases
with time as the probability distribution broadens; and, most importantly,
the width of $P_{m}(t)$ is proportional to $t$. This probability
distribution can be expressed, in some limit \cite{Knight}, as a
combination of Airy functions propagating in opposite directions.

The shape of $P_{m}(t)$ for $\alpha=0.4$ is very different: Now
the two peaks of $P_{m}(t)$ contain most of the total probability,
around 30\% each one in the case of Fig. 2, mostly distributed within
a few lattice positions (see the inset in Fig. 2) . But the most striking
characteristic of the probability peaks in this nonlinear case, is
that \textit{their size and shape remain basically constant with time},
except for small oscillations around a mean value.

We will characterize the probability peaks by their position and intensity
(i.e., the total probability they contain). As for the position, given
the small fluctuations on the shape of the peak, we use the \char`\"{}center
of mass\char`\"{}, defined as $m_{CM}\equiv\sum_{m}mP_{m}(t)$, with
$m\in\left[m_{\max}+\Delta m,m_{\max}-\Delta m\right]$, $m_{\max}$
the position of the probability maximum and $\Delta m$ the width
of the peak %
\footnote{We found that the choice $\Delta m=8$ is sufficient for our purposes.%
}. Only small quantitative differences are found between the behavior
of $m_{max}$ and that of $m_{CM}$.

\begin{figure}
\includegraphics[clip,scale=0.45]{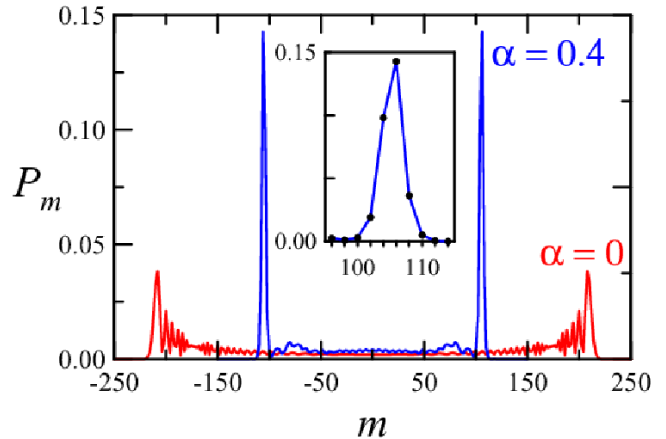}

\caption{(Color online) Probability distribution curves of $P_{m}(t)$ for
$t=300$, with the initial condition $u_{m,0}=\delta_{m0}/\sqrt{2}$
and $d_{m,0}=i\delta_{m0}/\sqrt{2}$, for $\alpha=0$ \emph{}(standard
QW) and $\alpha=0.4$. The inset is a magnification of the right--moving
probability soliton. Notice that $P_{m}\left(t\right)$ is null for
odd $m$ (as $t$ is even in this plot). We have represented only
nonzero values and joined them for guiding the eye.}
\end{figure}

The most important feature of the probability peaks is that they are
non-spreading pulses, i.e., they propagate without distortion %
\footnote{Except for the already mentioned temporal oscillations around a mean
value.%
}. As these probability wave-packets do not spread in time, and present
other particle--like features (see below) we can consider them as
solitonic-like structures, and will simply refer to them as solitons.

Apparently, solitons do not require a minimum value of $\alpha$ to
form: We have checked their existence for $\alpha\geq0.01$, and the
analysis of the data from different (non-zero) values of $\alpha$
does not suggest the existence of any threshold for the solitons formation.
Nevertheless, the time needed for their formation (i.e., the transient
until the intensity and shape of the probability structure is constant
on the average) is larger for smaller $\alpha$. This is appreciated
in Fig. 3 (a), where the intensity of a single soliton is represented
as a function of time for different values of the nonlinearity parameter
$\alpha$. Another important feature is that the width of the overall
probability distribution $P_{m}(t)$ or, equivalently, the soliton
velocity, decreases as $\alpha$ increases, as shown in Fig. 3 (b),
where the position of the solitons is represented for different values
of $\alpha$ (in the case $\alpha=0$, where solitons do not exist,
we have represented the position of the center of mass of the maximum
of $P_{m}\left(t\right)$ for the sake of comparison). Therefore,
solitons form after some transient, and are slower and more intense
for larger $\alpha$. This is the scenario we found for $\alpha\leq0.474$. 

\begin{figure}
\includegraphics[clip,scale=0.45]{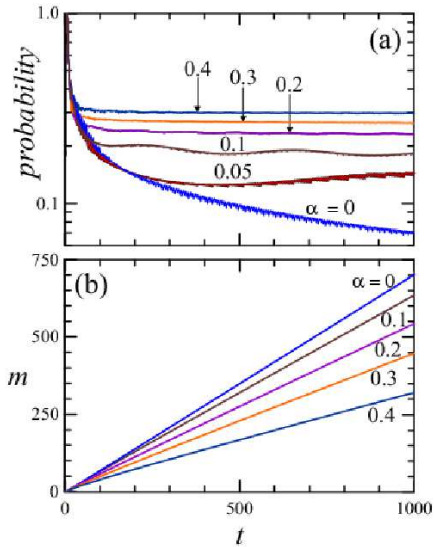}

\caption{(Color online) (a) Intensity (total probability) of the right--moving
soliton. (b) Temporal evolution of the center of mass, $m_{CM}$,
of the right--moving soliton (the plot is symmetric for the left-moving
soliton). The values of $\alpha$ are indicated in the plots. Initial
conditions are as in Fig. 2. }
\end{figure}

In view of the phenomena described above, one might wonder whether
the intrinsic quantum features of QWs are deteriorated or not, and
if so, to what extent. One way to quantify the possible loss of the
quantum benefits is by analyzing the time evolution of the standard
quadratic deviation $\sigma=\sqrt{<m²>-<m>²}$ . As already discussed,
the standard QW exhibits a characteristic $\sigma\propto t$. Given
the transient which appears during the formation of the solitons,
the question we ask ourselves is: does the quotient $\sigma/t$ go
to a constant after the transient (i.e., for sufficiently large time),
or will it decay slower? 

As can be seen from Fig. 4, the first possibility is in fact realized:
after the transient, the standard deviation approaches the typical
QW time evolution. Therefore, the long-term QW behavior is not degraded
by the formation and propagation of the solitons.

\textit{}%
\begin{figure}
\includegraphics[clip,scale=0.4]{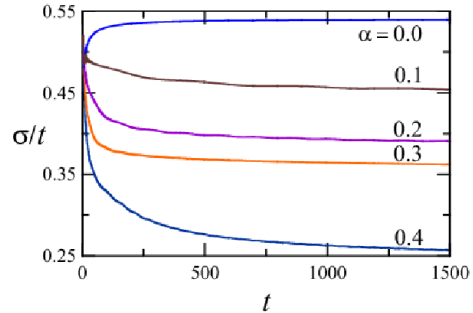}

\caption{(Color online) Evolution of the ratio $\sigma/t$ for different values
of $\alpha$. }
\end{figure}

\section{Dynamical phases.}

We have been able to identify three different dynamical domains, or
dynamical phases, in the behavior of solitons as a function of the
value of $\alpha$: Phase I, for $\alpha<\alpha_{I}\simeq0.474$;
phase II, for $\alpha_{I}<\alpha<\alpha_{II}\simeq0.6565$; and phase
III, for $\alpha>\alpha_{II}$. Let us describe these phases separately.

In Phase I, the dynamics is very simple: Once solitons have formed,
they exhibit the ballistic propagation already shown in Fig. 3(b).
Differently, in Phase II the two solitons start moving in opposite
directions, as in Phase I, but after some time their velocity decrease
till the solitons reach a \textit{turning point} and then move backwards
and collide at some later instant $t_{col}$ at $m=0$. After the
collision, the solitons continue moving apart indefinitely, as in
phase I. An example of such behavior, for $\alpha=0.49$, is shown
in Figure 5. Notice the appearance of small {}``communication packets''
that are interchanged between the two solitons. Interestingly, the
solitons intensity sharply decreases after the collision (for example,
for $\alpha=0.49$ the intensity of one soliton falls from $0.3062$
before the collision, to $0.2426$ afterwards), i.e., the collision
of the two solitons is an \textit{inelastic} one. Another feature
that can be observed from the simulations is that, as $\alpha$ is
increased from below $\alpha_{I}$ (inside phase I), the intensity
of the solitons increase up to a maximum value. It seems that the
communication packets interchanged by the two solitons play the role
of an attractive interaction, which is larger for larger intensities.
This would explain the existence of the above-mentioned turning point
appearing at some critical value $\alpha_{I}$. Inside phase II, the
solitons experience an inelastic scattering and loose a fraction of
their intensity, which would prevent from recollapse.

\begin{figure}
\includegraphics[clip,scale=0.45]{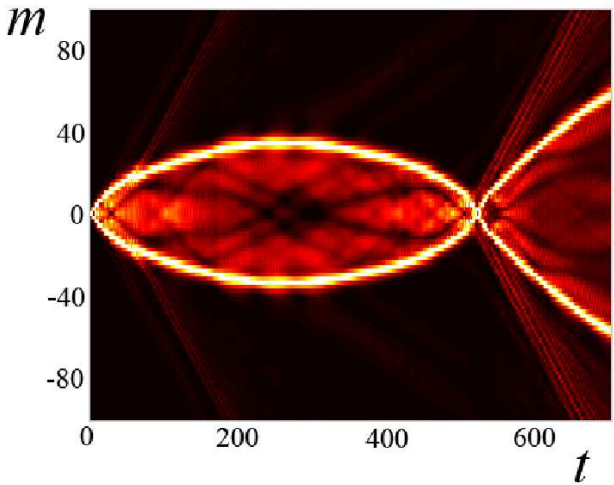}

\caption{(Color online) Color density plot showing the evolution of $P_{m}(t)$
as a function of $t$ (horizontal axis) for $\alpha=0.49$. The vertical
axis corresponds to the position on the lattice. Brighter regions
indicate a higher probability. The two solitons are clearly visualized
as intense strips.}
\end{figure}

The method we used to determine this critical value, however, makes
use of the fact that the collision instant $t_{col}$ decreases with
$\alpha$. Indeed, the function $t_{col}(\alpha)$ can be well reproduced
numerically by a simple hyperbola $1/t_{col}=a/\alpha+b$ (where the
values of $a$ and $b$ are obtained by a numerical fit, with a coefficient
of determination $r^{2}=0.99516$, giving $a=-0.0297\pm0.0003$ and
$b=0.0627\pm0.0006$). This numerically-obtained law allows to fix
the frontier between phases I and II by the $\alpha$ value for which
$t_{col}$ diverges (we obtained $\alpha_{I}=0.474\pm0.007$).

As we made for phase I, it is worth investigating how the standard
deviation evolves at long times, in order to quantify a possible departure
from the characteristic quantum spreading. As before, we plot in Fig.
6 the quotient $\sigma/t$ as a function of time for values of $\alpha$
corresponding to the second phase. Now the transient shows more complicated
features, due to the recollapse of the two solitons (which manifests
as the minimum appearing in both curves). However, as the solitons
separate after the collision, the typical $\sigma\propto t$ behavior
shows up.

\begin{figure}
\includegraphics[clip,scale=0.4]{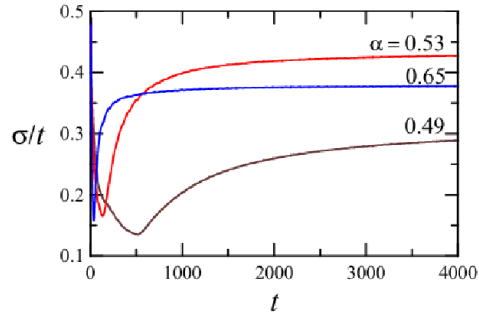}

\caption{(Color online) Same as Fig. 4, for values of $\alpha$ corresponding
to phase II.}
\end{figure}

Phase III, $\alpha>\alpha_{II}$, differs from phase II in that, after
the collision, the two emerging solitons do not necessarily collide
or separate from each other. In fact, if $\alpha$ is increased beyond
$\alpha_{II}$, the situation becomes quite complicated, as the evolution
of the solitons becomes extremely sensitive to small variations in
$\alpha$. In this sense, we can say that phase III is a chaotic phase:
For some values of $\alpha$, the solitons become trapped and oscillate
around the origin; with a slightly different value for $\alpha$,
however, the solitons eventually escape; and there are other $\alpha$
values for which localization is found, which is characterized by
an asymptotic setup of both solitonic structures at an equilibrium
point. Interestingly, the latter possibility can occur at very distant
site positions for slightly different values of $\alpha$: For example,
the right-moving soliton position oscillates between $m=5$ and $m=9$
for $\alpha=0.6665$; it remains static at position $m=162$ for $\alpha=0.6669$;
and again oscillates, around $m=7$, for $\alpha=0.6673$. In Fig.
7 we show an example of the type of dynamics one encounters in phase
III for the two values of $\alpha$ indicated in the figure caption.

\begin{figure}
\includegraphics[clip,scale=0.2]{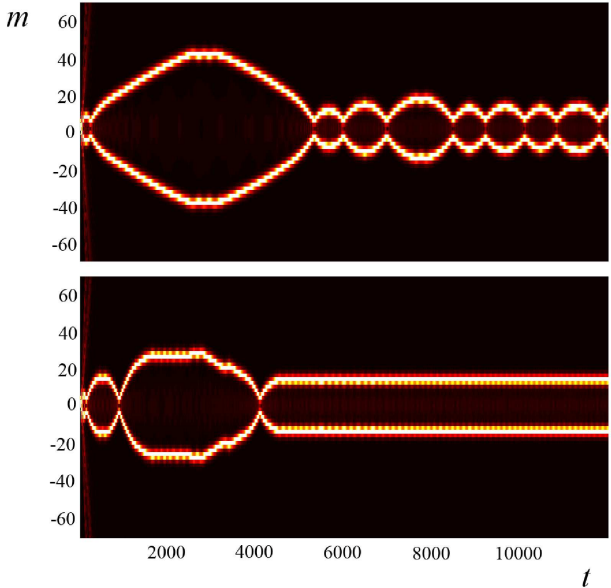}

\caption{(Color online) Same as Fig. 5, but for $\alpha=$$0.6565$ (top)
and $\alpha=0.658197$ (bottom).}
\end{figure}

The results we have just described correspond to a particular choice
of the QW initial conditions, which guarantees the symmetry of the
probability distribution with respect to the starting position. In
order to see how critical is the role of the initial condition, we
have carried out numerical simulations for different sets of initial
conditions, and have found that the dynamics is also very sensitive
to this choice. Fig. 8 gives an idea of how different things can be:
we represent the evolution of the probability distribution for $u_{m,0}=\delta_{m0},d_{m,0}=0$
and $\alpha=0.2$ (top) or $\alpha=0.6$ (bottom). For this initial
condition, the probability distribution is no longer symmetric (even
in the standard QW), and this fact strongly affects the formation
and dynamics of solitons. We are not going to enter into an exhaustive
description here; it will suffice to say that, in this case, there
are also several dynamic phases: For small $\alpha$, a single soliton
forms, carrying close to 60\% of the probability, that moves like
in Fig. 7 (top) (most of the rest of the probability is contained
in small dispersive pulses that can be appreciated in the figure);
for large $\alpha$ several solitons, with different intensities,
can form, and localization phenomena similar to what we have described
above can occur too, see Fig. 8 (bottom).

\begin{figure}
\includegraphics[clip,scale=0.2]{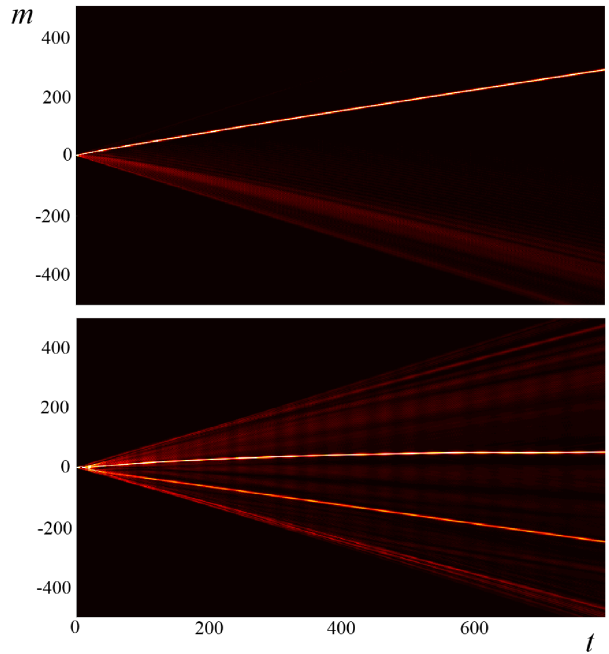}

\caption{(Color online) Same as Fig. 5, but for $\alpha=0.2$ (top) and $\alpha=0.6$
bottom. The initial conditions are $u_{m,0}=\delta_{m0}$ and $d_{m,0}=0$.
For $\alpha=0.2$ a single soliton is formed that carries 57\% of
the probability. For $\alpha=0.6$ this soliton has now a smaller
intensity (32\% of the probability) and becomes localized near $m=0$,
while a second soliton (20\% of the probability) is formed.}
\end{figure}

\section{Conclusions.}

We have introduced a simple variation of the Optical Galton Board
(which can be understood as a classical implementation of the discrete
coined QW), based on the assumption that light propagates through
a nonlinear (Kerr-type) medium inside the optical cavity or, using
the algebraic language of QW, based on the acquisition of non-linear
-probability dependent- phases by the state during the walk.

The most striking feature that the nonlinearity introduces, is the
formation of soliton-like structures, which carry a constant fraction
of the total intensity (probability) distribution within a non-dispersive
pulse. We have characterized the dynamics of these solitons showing
the existence of complex dynamics (from ballistic motion to dynamical
localization) that is very sensitive to the initial conditions. An
important feature we found is that, in spite of the complicated behavior
during the transient and possible recollapse of the solitons, the
long term evolution still shows the characteristic QW feature in the
cases when the solitons go away, in the sense that the standard deviation
becomes $\sigma\propto t$.

It would be of greatest interest to have at hand an analytical description
of the solitons motion and interaction, specially during the formation
transient and recollapse (when present), as done (approximately) in
{[}10]. The additional complication due to non-linearities, however,
makes this task cumbersome and lies beyond the scope of this paper.

The described phenomena are, to the best of our knowledge, new in
the field of quantum walks. The exciting features found here deserve,
we believe, further research.

\textbf{Acknowledgments}

We gratefully acknowledge fruitful discussions with G.J. de Valcárcel.
This work has been supported by the Spanish Ministerio de Educación
y Ciencia and the European Union FEDER through Projects FIS2005-07931-C03-01,
AYA2004-08067-C01, FPA2005-00711 and by Generalitat Valenciana (grant
GV05/264).


\begin{thebibliography}{10}
\bibitem{nota}We must emphasize that $P_{m}^{u}(t)$ and $P_{m}^{d}(t)$
are probabilities in a true quantum system, but in a classical system
they are intensities, see e.g. \cite{Knight}.
\end{thebibliography}
\end{document}